\documentstyle[aps,psfig]{revtex}

\title{Influence of correlations on the velocity statistics of scalar granular gases}

\author{A. Baldassarri$^1$, U.
  Marini Bettolo Marconi$^1$, A. Puglisi$^2$} \address{$^1$ INFM Udr Camerino,
  Univ. di Camerino, Dip. Matematica e Fisica, Via Madonna delle
  Carceri I-62032 Camerino, Italy} \address{$^2$ Universit\'a ``La
  Sapienza'', P.le A. Moro 2, 00185 Roma, Italy} 
\date{\today}

\begin{document}

\maketitle

\begin{abstract}
  The free evolution of inelastic particles in one dimension is
  studied by means of Molecular Dynamics (MD), of an inelastic
  pseudo-Maxwell model and of a lattice model, with emphasis on the
  role of spatial correlations. We present an exact solution of the 1d
  granular pseudo-Maxwell model for the scaling distribution of
  velocities and discuss how this model fails to describe correctly
  the homogeneous cooling stage of the 1d granular gas. Embedding the
  pseudo-Maxwell gas on a lattice (hence allowing for the onset of
  spatial correlations), we find a much better agreement
  with the MD simulations even in the inhomogeneous regime. This is
  seen by comparing the velocity distributions, the velocity profiles
  and the structure factors of the velocity field.
\end{abstract}

\section{Introduction}

Some twenty years ago, S.Ulam proposed an elegant and simple
model~\cite{ulam}
aimed to show how the Maxwell distribution for the velocity statistics
in a gas may be derived directly from the conservation laws encoded in the
particle dynamics. He considered the evolution of
an assembly of $N$ particles, whose
initial velocity distribution
is arbitrary. At each instant a random
pair of particles is selected and their velocities are updated 
as if they had collided
elastically. Such a process converges spontaneously to a macro-state
whose velocity distribution is
a Gaussian, with the variance given by the constant energy of the system.

Relaxing the energy conservation law, i.e. considering inelastic collisions
between particles, leads to different statistical properties, which
in fact are related to a well defined physical problem, i.e. 
%% Intro generale, il cooling (3 regimi), i modelli semplificati
the dynamics of Fluidized Granular Materials~\cite{review}. These are
assemblies of moving macroscopic particles, whose evolution is
controlled by the inelasticity of their mutual collisions.
They present unusual structural and statistical properties which
stimulated many studies not only because of
their industrial and technological relevance, but also due to the fundamental
questions they raise.
%% Il problema delle distribuzioni 
%%Even the functional form of the velocity density distribution function
%%of granular systems may depart from the ubiquituous Maxwell
%%distribution, which characterizes molecuar fluids.  

Roughly speaking, the inelasticity causes two phenomena on the
velocity statistics: a) in the absence of energy injection the fluid cools 
down since the variance of the velocity distribution, the so called granular
temperature, decreases; b)
the form of the distribution may even cease to be Gaussian and
approach a stationary shape under a suitable rescaling.  
%%Are there
%%universal feature in the shapes of the distribution for systems with
%%different inelasticity and dimensionality?

A large amount of recent studies~\cite{qualcosa} indicates that the
evolution of a freely evolving granular system consists of three
stages: a homogeneous cooling during which both the velocity and the
density are spatially uniform; a second stage where the velocity field
forms vortices and a third stage where the density becomes highly
inhomogeneous and clusters appear.  Unfortunately, many of the
existing approaches have intrinsic limitations: MD numerical
simulations are time demanding and hardly access to the correlated
regimes; kinetic theories, based on the molecular chaos assumption,
seem to be limited to the quasi elastic situation; hydrodynamic
equations suffer from their phenomenological derivation. For these
reason minimalistic models, which either lend themselves to analytical
solutions or greatly reduce the computational effort, can provide
useful hints.

Hereafter we shall focus on the physics of rapid granular flows 
employing a one dimensional collision rule in order to 
render the analysis as simple as possible. This choice,
in spite of its apparent simplicity shares many
features with the physics of higher dimensions. 
The paper is organised as follows: first, we shall
revisit the inelastic hard rod model moving on a ring 
by means of event driven dynamics. 
Next, we shall compare these
results with those of the inelastic Ulam
model (which corresponds to a gas of pseudo-Maxwell
molecules~\cite{krapivsky,bobylev}),  for which we find the exact 
asymptotic velocity statistics, showing how they disagree even in
the homogeneous regime. Thirdly, we shall extend Ulam's
model including spatial fluctuations, and show that 
many features of the real one-dimensional gas, 
both in the homogeneous and in the inhomogeneous
regime, are recovered.

\section{Inelastic hard rods on a ring}
%% Modelli unidimensionali

We define a one dimensional granular gas as a system
of hard rods, confined to a ring and colliding
inelastically. After a binary collision the scalar velocities of the
particles change according to:  

\begin{eqnarray}
 v_i' & = & v_i-
\frac{ 1+r}{2}(v_i-v_j) \\
 v_j' & = & v_j+
\frac{ 1+r}{2}(v_i-v_j)
\label{collision}
\end{eqnarray}

where $r$ is  a restitution coefficient, which takes 
on the value $1$
for perfectly elastic systems and $0$ for completely inelastic
particles.
Few years ago McNamara and Young~\cite{mcnamara1}, and
Sela and Goldhirsh~\cite{sela}, simulated such a 
model and observed a universal algebraic decay of the kinetic energy,
$E(t)\propto t^{-2}$ (Haff's Law), together with an anomalous
behavior for the global velocity distribution, even in the early homogeneous
regime. Furthermore, the appearance of strong inhomogeneities
is the precursor of a numerically catastrophic event, 
named  inelastic collapse: i.e. particles perform an infinite number of
collisions in finite time interval.
%%  1d MD (Ben Naim, Redner) e limite quasi elastico della Boltzmann (Caglioti)
A renewal of interest on one-dimensional granular flows, has been
generated by the recent work of Ben-Naim et al.~\cite{redner}. They
eliminated the inconvenience of the inelastic collapse, by means of a
physically motivated 'regularization': they assumed the restitution
coefficient to be elastic for small momentum transfer.  This allows
the system to enter into a new dynamical regime, independent of the
choice of the regularization.  During such a regime, the kinetic energy decays
as $E(t)\propto t^{-2/3}$, for any $r<1$. A
direct inspection of the hydrodinamic profiles, shows that such
a regime is highly inhomogeneous, with density clusters and shocks in
the velocity field.  They suggest that inelastic systems
behave asymptotically as a sticky ($r=0$) gas~\cite{pomeau}, which 
is known to be described by the Burgers equation 
in the inviscid limit~\cite{zeldovich}. This should justify the
presence of shocks and  clustering and, moreover, predicts
that the tails of the velocity distribution decay as
$\exp(-v^3)$~\cite{frachebourg}. On the other hand,
an accurate numerical test of this prediction~\cite{redner}
seems problematic, because 
the bulk of the distribution does not deviate appreciably 
from a Gaussian (see Fig.~\ref{fig1b}).
Such a behavior reflects the fact
that the asymptotics is dominated by the dynamics of
clusters of particles, which move through the system and coalesce,
like sticky objects. 
Although the sticky nature of the inelastic granular dynamics seems a
very appealing description, it is not clear if it applies to higher
dimensions~\cite{no-burgers}.
%% since there is no rigorous analytic derivation of the 
%% above connection. 

%%%%%%%%%%%%%%%%%%%%%%%%%%%%%%%%%%%%%%%%%%%%%%%%%%% FIG 1
%
% Distributions for MD and lattice model
%
%%%%%%%%%%%%%%%%%%%%%%%%%%%%%%%%%%%%%%%%%%%%%%%%%%%
\begin{figure}[t]
\centerline{ \psfig{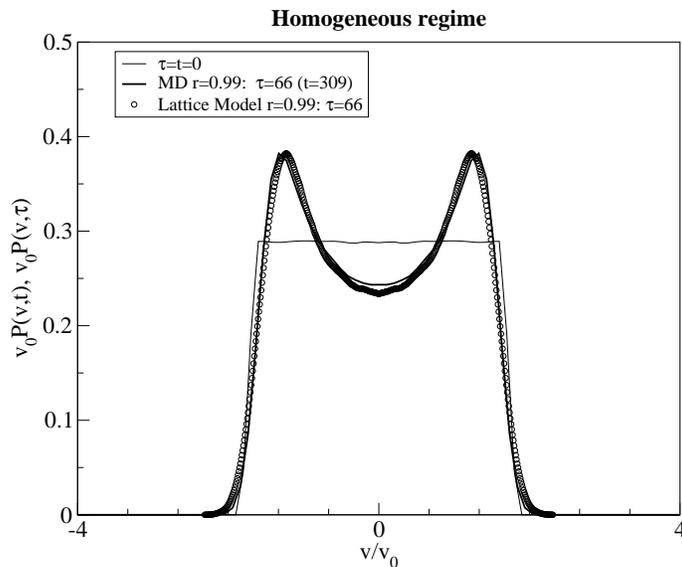} }
\caption
{ Rescaled velocity distributions for
  the MD and in the lattice gas, during the
  homogeneous. The initial distribution (both
  models) is also shown.  The distributions refer to systems having the same
  energy. Data refer to $N=10^6$ (both models) particles with $r=0.99$ and $r=0.5$
  (for the lattice model in the inhomogeneous regime).
}
\label{fig1a}
\end{figure}

To the best of our knowledge, exact analytic treatments of the hard rod model
are limited to the homogeneous regime, by means of solutions
of the Boltzmann equation. Caglioti et
al.\cite{caglioti} solved asymptotically the Boltzmann equation for a 1d
inelastic hard rod gas in the limit $r \to 1^-$ and $\rho \to \infty$
(where $\rho$ is the density). They obtained the asymptotic velocity
distribution, which, if rescaled to have a constant variance, is 
the superposition of two delta functions (see also~\cite{Barrat}). 
For MD simulations, the
observation of two distinct peaks in the rescaled velocity
distribution was reported in~\cite{sela} (see
Fig.~\ref{fig1a}): it represents a precursor
of that singular asymptotic distribution. However, the appearance of
inhomogeneities and, eventually, the inelastic collapse prevent the
approach to the limiting distribution.

%%%%%%%%%%%%%%%%%%%%%%%%%%%%%%%%%%%%%%%%%%%%%%%%%%% FIG 1
%
% Distributions for MD and lattice model
%
%%%%%%%%%%%%%%%%%%%%%%%%%%%%%%%%%%%%%%%%%%%%%%%%%%%
\begin{figure}[t]
\centerline{ \psfig{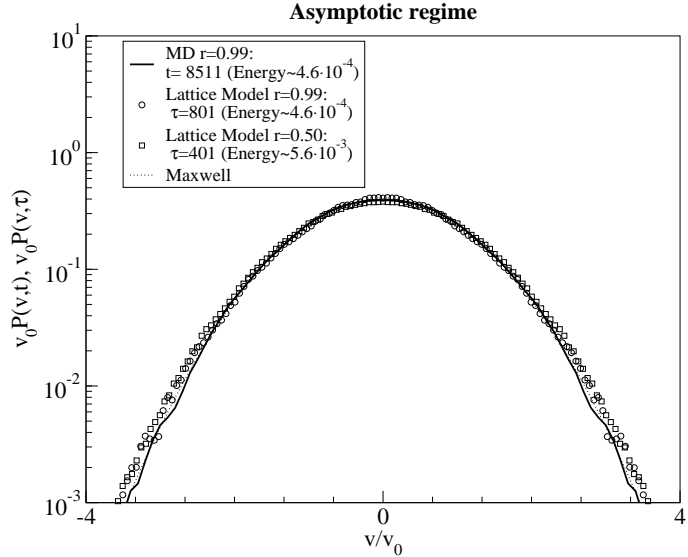} }
\caption
{ Rescaled velocity distributions for
  the 1D MD and in the 1D lattice gas, during the
  inhomogeneous phase. The distributions refer to systems having the same
  energy. Data refer to $N=10^6$ (both models) particles with $r=0.99$ and $r=0.5$
  (for the lattice model in the inhomogeneous regime).
}
\label{fig1b}
\end{figure}                                   

%%%%%%%%%%%%%%%%%%%%%%%%%%%%%%%%%%%%%%%%%%%%%%%%%%% FIG 1
%
%  Lattice vs. MD
% 
%%%%%%%%%%%%%%%%%%%%%%%%%%%%%%%%%%%%%%%%%%%%%%%%%%%
\begin{figure}[tbp]
\centerline{ \psfig{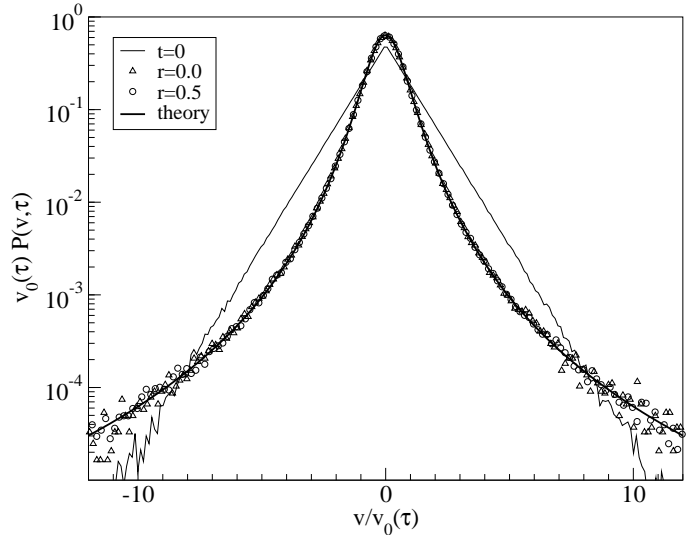} }
\caption
{Asymptotic velocity distributions $P(v,\tau)$ versus $v/v_0(\tau)$
  for different values of $r$ from the simulation of the inelastic
  pseudo-Maxwell (Ulam's) model. The asymptotic distribution is
  independent of $r$ and collapse to the
  Eq.(\protect{\ref{exact_v}}). The chosen initial distribution is
  drawn (same result with uniform and Gaussian initial distribution).
%   In 2D the distributions still present power-law tails, but the power
%   depends upon $r$, as shown in the inset: for $r=0$ data are
%   compatible with $\alpha=5$, while for larger $r$, $\alpha$ increases
%   and for $r \to 1$ tends to a Maxwell distribution. Data refers to
  more than $N=10^6$ particles.
}

\label{fig2}
\end{figure}                                   

\section{Ulam's model}
%% Modelli a la Ulam
Alternatively, one may attack the Boltzmann equation, by 
assuming a simpler form for
the scattering cross section in the collision integral, i.e.
taking 
the relative velocity of the colliding pair to be proportional to the
average thermal velocity: $|v-v'| \sim \sqrt{E}$. This is the 
so called pseudo-Maxwell model \cite{bobylev}. The energy
factor  $\sqrt{E}$ can be eliminated via a time reparametrization,
and one obtains a simpler equation:
\begin{equation}
%%\partial_t P(v,t)+P(v,t)= \frac{1}{1-\gamma}\int
%%duP(u,t)P\left(\frac{v-\gamma u}{1-\gamma},t\right)
%%\partial_t P(v,t)+P(v,t)={1\over 1-\gamma}
%%\int du P(u,t)P\left({v-\gamma u\over 1-\gamma},t\right)
\partial_\tau P(v,\tau)\!+\!P(v,\tau)\!=\!\beta\!\!\int\!\!\!du\, 
P(u,\tau)P\left(\beta v\!+\!(1\!-\!\beta) u,\tau\right)
\label{ulam1d_v}
\end{equation}
where $\beta=2/(1+r)$ and the $\tau$ counts the number of collisions
per particle.  Eq. (\ref{ulam1d_v}) is the master equation of the
inelastic version of Ulam's scalar model: at each step an arbitrary
pair is selected and the scalar velocities are transformed according
to the rule of Eq. (\ref{collision}). In a stimulating paper, Ben-Naim
and Krapivsky~\cite{krapivsky} considered such scalar model, and
obtained the evolution of the moments of the velocity
distributions. Since at large times, $\left<v^n\right>\sim \exp(-\tau
a_n)$, and the decay rates $a_n\neq n a_2/2$ (they depend non-linearly
on $n$), they argued that such a multiscaling behavior prevents the
existence of a rescaled asymptotic distribution $f$ such that
$P(v,\tau)\to f(v/v_0(\tau))/v_0(\tau)$, for large $\tau$, where
$v_0^2(\tau)=\int v^2 P(v,\tau)dv=E(\tau)$. On the contrary, we
believe that such ``multiscaling'' behavior only indicates the fact
that the moments of the rescaled distribution $\int x^n f(x) dx=\left<
v^n\right>/v_0^n$ diverge asymptotically for $n\ge 3$, and does not
rule out the possibility of the existence of an asymptotic
distribution with power law tails.  In fact, the Fourier transform of
Eq.~(\ref{ulam1d_v})

%%%%%%%%%%%%%%%%%%%%%%%%%%%%%%%%%%%%%%%%%%%%%%%%%%% FIG 3
%
% Structure factors
%
%%%%%%%%%%%%%%%%%%%%%%%%%%%%%%%%%%%%%%%%%%%%%%%%%%%
\begin{figure}[tbp]
\centerline{ \psfig{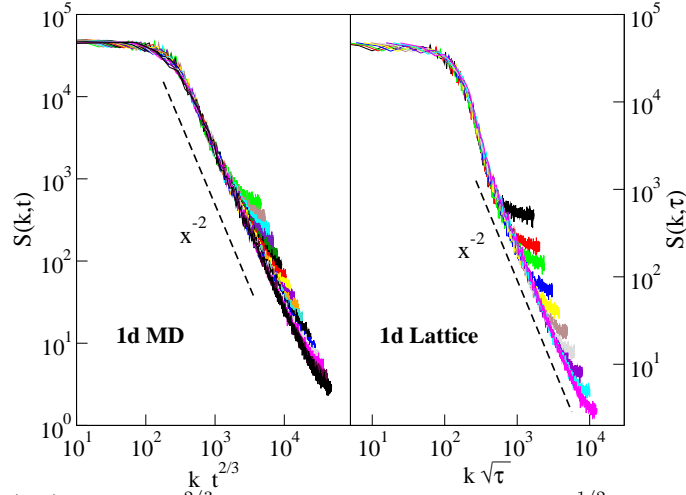} }
\caption
{Structure factors $S(k,t)$ against $kt^{2/3}$ for the 1D MD 
 and against $k\tau^{1/2}$ for the 1D lattice gas model, in
 the inhomogeneous phase. Times are chosen so that the two
  systems have the same energies. Data refers to system with more than
 $N=10^5$ particles, $r=0.5$ (both models).}
\label{fig3}
\end{figure}                                   

\begin{equation}
\partial_\tau
\hat{P}(k,\tau)+\hat{P}(k,\tau)=\hat{P}[ k/(1-\beta),\tau]\hat{P}[k/\beta,\tau]
\label{ulam1d}
\end{equation}
possesses several self-similar solutions of the kind $\hat{P}(k,\tau)=\hat
f(k v_0(\tau))$, which correspond to the asymptotic rescaled distribution
$P(v,\tau)=f(v/v_0(\tau))/v_0(\tau)$. Many of them do
not correspond to physically acceptable velocity
distributions~\cite{krapivsky}. 
The
divergence of the higher moments implies a non analytic
structure of $\hat f$ in $k=0$, since $\left<v^n\right>/v_0^n=(-i)^n
\frac{d^n}{dk^n} \hat{f}(k)|_{k=0}$, and 
represents a guide in 
the selection of the physical solution.
As shown in Fig.~\ref{fig2}, our data 
collapse on the function
\begin{equation}
f(v/v_0(\tau))=\frac{2 }{\pi \left[1+(v/v_0(\tau))^2\right]^2}.
\label{exact_v}
\end{equation}
corresponding to the self-similar solution
$\hat{f}(k)=(1+|k|)\exp(-|k|)$~\cite{note}.  Notice that
(\ref{exact_v}) is a solution of Eq.(\ref{ulam1d}) for every $r<1$,
i.e.  the asymptotic velocity distribution does not depend on the
value of $r<1$, as shown in Fig.~\ref{fig2} (left frame).  However,
such solution~(\ref{exact_v}) does not resemble the distribution
obtained in the simulation of the inelastic hard rod model, suggesting
that the approach {\em \'a la Ulam} fails to describe even the
homogeneous cooling phase (see Fig.~\ref{fig1a}).  We believe that
such failure is due to the absence of velocity correlations.

\section{One-dimensional lattice model}
 
To reinstate spatial correlations we considered a 1d lattice model
where the $N$ sites have a ``velocity'' $v_i$. At every step of the
evolution a pair of neighboring sites is chosen randomly and undergoes
an inelastic collision according to Eq.(\ref{collision}) if
$(v_i-v_j)\times (i-j)<0$. The latter condition, which avoids
collisions between particles ``moving'' away from each other, is to be
be referred below as the {\em kinematic constraint} and plays a key
role in the formation of structures during the inhomogeneous phase. A
unit of time $\tau$ correspond to $N$ of such steps (i.e. to $1$
collision per particle on average). Note that, since there is no real
particle movement, the number of collisions per particle, $\tau$,
remains the only measure of time in our model.  This represents a
problem in the direct comparison with MD, because $\tau(t)$ is a well
defined ``universal'' function for the homogeneous regime only: in the
inhomogeneous regime $\tau(t)$ diverges due to the inelastic collapse,
or depends on the ``regularization'' of the restitution
coefficient. We observe that initially the energy $E(\tau)$ decreases
exponentially, as in the Haff regime for MD simulation, but after the
formation of long range velocity correlations, the decay slows down,
and a power law decay of the form $E \sim
\tau^{-1/2}$, appears.
As stated above, this law cannot 
be directly compared to the $t^{-2/3}$ behavior,  observed in MD,
since, in the non-homogeneous regime the function $\tau(t)$ depends on
the
regularization employed to avoid the inelastic collapse.
However the problem can be circumvented if one {\em assumes} the
energy, instead of $\tau$, as the physical clock of both dynamics.

\begin{figure}[tbp]
\centerline{ \psfig{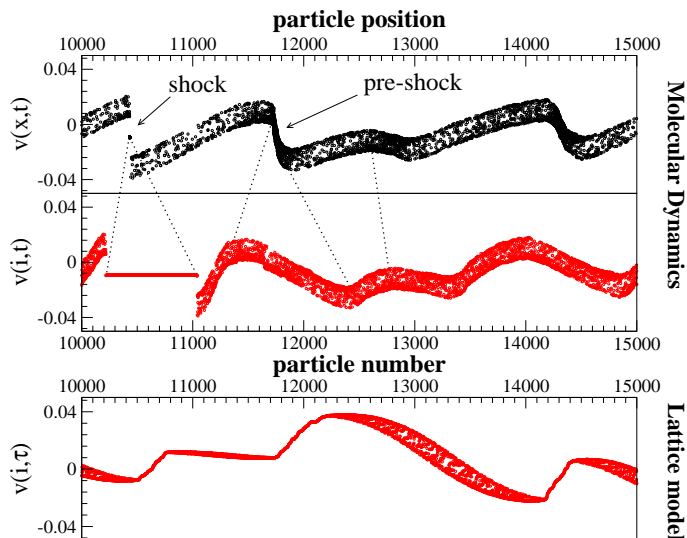} }
\caption
{Portions of the instantaneous velocity profiles for the 1D MD
  (top $v(x,t)$, middle $v(i,t)$) and for the 1D lattice
  gas model (bottom, $v(i,\tau)$). In the middle frame we display the
  MD profile against the particle label in order to
  compare the shocks and preshocks structures with the lattice gas
  model (the dotted lines show how shocks and preshocks transform
  in the two representations for the MD). Data refers to $N=2\cdot
  10^4$ particles, $r=0.99$ (both models). 
  }
\label{fig4}
\end{figure}

A surprising agreement between the MD and the lattice gas is found in
the comparison of the velocity distribution taken at instants
corresponding to the same energy.
% Figs. compare the velocity distributions 
% of the MD and of the lattice gas respectively at an early and at a late instant
% chosen in such a way that the MD and the lattice gas have
% the same energy.  
Both the early two humps structure and the late Gaussian shape seen in
MD simulations agree with those of the lattice gas, as shown in
Figs.~\ref{fig1a} and~\ref{fig1b}. The Gaussianity of the asymptotic
field distribution, and the $\tau^{-1/2}$ asymptotic decay suggest
that, in the lattice model, the scalar velocity field $v(i,\tau)$ is
governed by a discretized version of the diffusion equation of the
form: $\partial_\tau v(i,\tau)=\nu \Delta v(i,\tau)$, where $\Delta$
is a lattice laplacian.  We checked this possibility studying the
structure factor $S(k,\tau)=\hat{v}(k,\tau)\hat{v}(-k,\tau)$ where
$\hat{v}(k,\tau)$ is the Fourier transform of the field $v(i,\tau)$.
In fact, at late times, $S(k,\tau)=f_1(k\tau^{1/2})$ scales similarly
to a diffusive process as shown by the data collapse in
Fig.~\ref{fig3} (right frame). However the form of the scaling
function differs from the Gaussian shape predicted by the diffusion
equation.  In same figure (left) we report the structure factor of the
MD simulation (fourier transform of
$C(r,t)=\left<v(i,t)v(i+r,t)\right>$, where $v(i,t)$ is the velocity
of the $i$-th particle, see below). In this case a good collapse,
$S(k,t)=f_2(kt^{2/3})$, has been obtained~\cite{md-scaling}. The two
collapses can be directly compared, because, in both models, the
structure factor is a function of $k/E$, consistently with the
previous discussion on $t$ and $\tau$. The algebraic tails of the
structure factors observed in both models carry important information
about the nature of the growing velocity correlations. A comparison of
velocity profiles at instants having the same energy in the two models
is presented in Fig.~\ref{fig4}. The top frame displays $v(x,t)$ of
the MD, with the characteristic shock and preshock structures. In the
middle frame, it is shown the profile $v(i,t)\equiv v(x_i(t))$ of the
MD, where $i$ is the particle label. The difference between $v(x,t)$
and $v(i,t)$ in MD is in the sign of large gradients: $v(x,t)$
displays large negative gradients, while $v(i,t)$ has large positive
gradients. Finally, the bottom frame displays the $v(i,\tau)$ profile
of the lattice gas which fairly compares with the analogous MD
quantity.  The power law behavior $S(k) \sim k^{-2}$ observed in both
models are due to the presence of short wavelength defects,
viz. shocks, as predicted by Porod's law~\cite{bray}.  Note that the
presence of shocks in the lattice model is due to the kinematic
constraint, since they disappear, together with the Porod's tails,
relaxing the constraint.

%%%%%%%%%%%%%%%%%%%%%%%%%%%%%%%%%%%%%%%%%%%%%%%%%%% FIG 4
%
% Velocity profiles
%
%%%%%%%%%%%%%%%%%%%%%%%%%%%%%%%%%%%%%%%%%%%%%%%%%%%

\section{Conclusions}        

In summary, we have investigated the velocity statistics of the
homogeneous and  inhomogeneous cooling regimes of the 1d granular gas. 
The complex observed behavior, featuring two distinct dynamical
regimes, the onset of long-ranged velocity correlations and shocks in
the velocity profiles can be reproduced even disregarding
the real kinematics of the particles. In fact we introduced a lattice
model of immobile particles which reproduces quantitavely the velocity
statistics both at short and long times. The structure factors and the
velocity profiles can also be qualitatively compared.
The crucial ingredient of the model is the
onset of velocity correlations in the system. Such correlations are
relevant for the velocity statistics even when short-ranged. 
We corroborated such a conclusion through the study of the mean-field
limit of the lattice model, which corresponds to a pseudo-Maxwell
model of an inelastic gas. In this limit the velocity distributions
change drammatically, displaying large tails, at odds with
the real gas. We found the exact expression for this power law tailed 
distribution, which represent a self-similar solution of a non-trivial
integro-differential Boltzmann like equation~\cite{maxwell}.
Although the one-dimensional character of our results may seems
restrictive, they can be interesting in several contests, as 
kinetic theory and microscopic models for turbulence (Burgers
equation) or for surface growth (KPZ equation).

%%
%% Two dimensions
%%
%%

% Before concluding we shall comment about higher dimensions. We recall
% that an analytic study  of the pseudo-Maxwell 
% model for $d=3$ has been performed by 
% Bobylev et al\cite{bobylev}. Unfortunately their
% technique relies on a Taylor expansion near $k=0$, i.e. it implies an 
% analytic structure and does not allow for the emergence of power law tails. 
% On the other hand we
% performed simulations for $d=2$, and  discovered a
% stable scaling solution with power law tails $\propto
% (v/v_0)^{-\alpha}$, for $v\gg v_0$. 
% At odds with $d=1$,  the exponent $\alpha$ seems to  depend on the
% restitution coefficient (see Fig.~\ref{fig2}, right frame). 
% MD simulations for the homogeneous cooling of hard
% spheres, indicate, however, that the tails of the velocity
% distribution are much faster (they seems to agree with the 
% exponential decay predicted by the Boltzmann
% equation for $d\ge 2$ \cite{vannoije}). The lattice correction of the
% Ulam's model for $d=2$, which seems to recover the correct
% distributions,  has been presented elsewhere~\cite{noilatticegas}.

We wish to thank E. Caglioti and A. Vulpiani for
many fruitful discussions.


\begin{references}
\bibitem{note} This solution  is similar to a solution explicitly 
mentioned by in~\cite{krapivsky}, but for the appearance of the modulus
which determines the singularities in $k=0$.
\end{references}
\end{document}